\documentclass[twocolumn]{aastex63}

\usepackage{rotating}
\usepackage{booktabs}
\usepackage{dcolumn}
\usepackage{ulem}
\usepackage[flushleft]{threeparttable}

\received{February 19, 2020}
\accepted{April 20, 2020}
\submitjournal{ApJ}

\shorttitle{How activity alters stellar spectra}
\shortauthors{Spina et al.}

\begin{document}

\title{How magnetic activity alters what we learn from stellar spectra}

\correspondingauthor{Lorenzo Spina}
\email{lorenzo.spina@monash.edu}

\author[0000-0002-9760-6249]{Lorenzo Spina}
\affil{School of Physics and Astronomy, Monash University, VIC 3800, Australia}
\affil{ARC Centre of Excellence for All Sky Astrophysics in Three Dimensions (ASTRO-3D)}

\author{Thomas Nordlander}
\affil{Research School of Astronomy and Astrophysics, The Australian National University, Canberra, ACT 2611, Australia}
\affil{ARC Centre of Excellence for All Sky Astrophysics in Three Dimensions (ASTRO-3D)}

\author{Andrew R. Casey}
\affil{School of Physics and Astronomy, Monash University, VIC 3800, Australia}
\affil{ARC Centre of Excellence for All Sky Astrophysics in Three Dimensions (ASTRO-3D)}

\author{Megan Bedell}
\affil{Flatiron Institute, Simons Foundation, 162 Fifth Ave, New York, NY 10010, USA}

\author{Valentina D'Orazi}
\affil{INAF - Astronomical Observatory of Padua, vicolo dell-Osservatorio 5, 35122, Padova, Italy}

\author{Jorge Mel\'{e}ndez}
\affil{Universidade de S\~ao Paulo, IAG, Departamento de Astronomia, Rua do Mat\~ao 1226, S\~ao Paulo, 05509-900 SP, Brasil}

\author{Amanda I. Karakas}
\affil{School of Physics and Astronomy, Monash University, VIC 3800, Australia}
\affil{ARC Centre of Excellence for All Sky Astrophysics in Three Dimensions (ASTRO-3D)}

\author{Silvano Desidera}
\affil{INAF - Astronomical Observatory of Padua, vicolo dell'Osservatorio 5, 35122, Padova, Italy}

\author{Martina Baratella}
\affil{Department of Physics and Astronomy $Galileo$ $Galilei$, University of Padua, Vicolo Osservatorio 3, I-35122, Padova, Italy}
\affil{INAF - Astronomical Observatory of Padua, vicolo dell'Osservatorio 5, 35122, Padova, Italy}

\author{Jhon J. Yana Galarza}
\affil{Universidade de S\~ao Paulo, IAG, Departamento de Astronomia, Rua do Mat\~ao 1226, S\~ao Paulo, 05509-900 SP, Brasil}

\author{Giada Casali}
\affil{School of Physics and Astronomy, Monash University, VIC 3800, Australia}
\affil{INAF - Arcetri Astrophysical Observatory, Largo E. Fermi, 5, I-50125 Firenze, Italy}
\affil{Department of Physics and Astronomy, University of FLorence, via G. Sansone 1, 50019 Sesto Fiorentino (Firenze), Italy}



\begin{abstract}
Magnetic fields and stellar spots can alter the equivalent widths of absorption lines in stellar spectra, varying during the activity cycle. This also influences the information that we derive through spectroscopic analysis. In this study we analyse high-resolution spectra of 211 Sun-like stars observed at different phases of their activity cycles, in order to investigate how stellar activity affects the spectroscopic determination of stellar parameters and chemical abundances. We observe that equivalent widths of lines can increase as a function of the activity index log~R$^\prime_{\rm HK}$ during the stellar cycle, which also produces an artificial growth of the stellar microturbulence and a decrease in effective temperature and metallicity. This effect is visible for stars with activity indexes log~R$^\prime_{\rm HK}$$\geq$$-$5.0 (i.e., younger than 4-5 Gyr) and it is more significant at higher activity levels. These results have fundamental implications on several topics in astrophysics that are discussed in the paper, including stellar nucleosynthesis, chemical tagging, the study of Galactic chemical evolution, chemically anomalous stars, the structure of the Milky Way disk, stellar formation rates, photoevaporation of circumstellar disks, and planet hunting.

\end{abstract}




\section{Introduction} \label{sec:intro}
How does chromospheric activity affect the way we interpret stellar spectra? In recent years, both observational and theoretical studies have addressed this fundamental question for stellar astrophysics. The spectroscopic analysis of the Sun-like star HD~45184 performed by \citet{Flores16} revealed that the Fe~II lines at 4924 $\rm \AA$, 5018 $\rm \AA$, and 5169 $\rm \AA$, formed in the upper photosphere, have their equivalent widths (EWs) modulate over the stellar activity cycle. More recently, \citet{Galarza19} showed that the EWs of iron lines in the spectra of the young ($\sim$400 Myr) solar twin HD~59967 increase as a function of  chromospheric activity along the stellar cycle. They also demonstrated that the EW variations occur for quantities which depend on the mean line-centre optical depth of formation ($\tau_\lambda$). The direct consequence of this effect is an increase in atmospheric microturbulence ($\xi$) inferred from the relation between derived Fe abundances and reduced EW, which is proportional to stellar activity level. This effect also drove an artificial decrease of the stellar metallicity ([Fe/H]) and effective temperature (T$_{\rm eff}$) as a function of chromospheric activity. No variations were observed for surface gravity (log~g). 

The results from \citet{Flores16} and \citet{Galarza19} confirm and conclusively demonstrate the hypothesis advanced by other observational studies, that elemental abundances in stellar spectra can correlate with stellar activity (e.g., \citealt{Morel03,Morel04,Reddy17,Baratella20}). 

These observations are also supported by theoretical works showing that the presence of a magnetic field can affect spectral lines, both directly through the Zeeman effect and indirectly, due to the magnetically induced changes on the thermodynamical structure of the atmosphere (e.g., \citealt{Borrero08,Fabbian10,Fabbian12,Moore15,Shchukina15,Shchukina16}). Since the strength of magnetic fields in the stellar atmosphere changes following the activity cycle \citep{Babcock59}, this could explain the observed EW modulation as a function of the activity level. Finally, also the fraction of stellar surface covered by cool spots changes during the activity cycle \citep{Schwabe44}, which may play a role in varying the EWs, especially those from lines with low excitation potentials.

In spite of that, magnetic fields and cool starspots are usually neglected in the analysis of stellar spectra, on the unproven assumption that their effects are of secondary importance compared to other sources of uncertainty. Therefore, studying the effects of magnetic activity on stellar spectra is clearly an important new step forward in the progress of techniques for spectroscopic analysis.

With the present study we aim at extending the experiment performed on a single star by \citet{Galarza19} to 211 Sun-like stars observed 21,897 times by the high-resolution spectrograph HARPS at different phases along their activity cycle (see Section~\ref{sec:analysis} for a detailed discussion on the spectroscopic analysis). Our final goal is to establish - over a large sample of stars covering a wide range of activity levels - how chromospheric activity can indirectly affect the absorption lines of stellar spectra and the information that we infer from spectroscopic analyses (see Section~\ref{sec:Results}). Our results have fundamental implications for several topics in astrophysics that are discussed in Section~\ref{sec:puzzles} and include stellar nucleosynthesis, chemical tagging, the study of Galactic chemical evolution, chemically anomalous stars, the structure of the Milky Way disk, stellar formation rates, photoevaporation of circumstellar disks, and planet hunting. Finally, in Section~\ref{sec:conclusions} we summarise the outcomes of this experiment and we draw our conclusions.

\section{Spectroscopic analysis} \label{sec:analysis}

Our experiment is carried out over a sample of stellar spectra collected by the HARPS spectrograph \citep{mayor03} and stored in the ESO Archive. The HARPS spectrograph is installed on the 3.6 m telescope at the ESO La Silla Observatory (Chile) and delivers a resolving power of 115,000 over the 383 - 690 nm wavelength range. The stars and spectra employed in our analysis are selected through the following criteria.

\begin{itemize}
\item We select the stars observed by HARPS with parameters falling within the following intervals: T$_{\rm eff}\in$[5500, 6100] K, log~g$\in$[4.0, 4.8] dex, and [Fe/H]$\in$[-0.3, 0.3] dex. The stellar parameters are obtained by Casali et al. (submitted) through the analysis of the co-added HARPS spectrum of each target using the line-by-line differential technique relative to the Solar spectrum. This technique has been developed (e.g.,\citealt{Langer98,Gratton01,Laws01,Melendez09}) to obtain precise differential abundances of similar stars, such as binary stars with similar components (e.g. \citealt{Desidera04,Ramirez11,Liu2014,Biazzo15,Teske2016,Nagar19}) and solar twin stars (e.g., \citealt{Ramirez09,Bedell14,Nissen15,Spina18b}).
\item Stars with at least 10 HARPS spectra available from the ESO public archive with signal-to-noise ratio S/N $\geq$ 100~pixel$^{-1}$ and acquired at airmass $\leq$1.6.
\item Stars with an intrinsic variation of chromospheric activity of $\Delta$log R$^\prime_{\rm HK} \geq$ 0.015 dex measured over the HARPS spectra\footnote{The log R$^\prime_{\rm HK}$ index measures the stellar chromospheric flux emission from the photospheric emission in the core of Ca H and K lines \citep{Noyes84}.}. The log~R$^\prime_{\rm HK}$ values are obtained through Eq. 6, 7, and 8 in \citet{Lorenzo-Oliveira18} and the measure on each exposure of the Ca II H$\&$K activity indices S$_{\rm HK}$ are performed according to the methods of \citet{Lovis11}. 
\end{itemize}

Our final sample includes 211 stars observed by HARPS 21,897 times in total. In Table~\ref{spectra} we list the ID of each spectrum, the corresponding star, the ESO project ID, the S/N measured on the 65$^{\rm th}$ spectral order, the airmass, exposure time and the barycentric Julian date (BJD) of the observation. Before the analysis, all spectra are normalised and Doppler-shifted using IRAF's \texttt{continuum} and \texttt{dopcor} tasks.

For the spectroscopic analysis, we employed a line list consisting of 78 Fe I lines, 17 Fe II lines and 146 lines of other elements (i.e., C, Na, Mg, Al, Si, S, Ca, Sc, Ti, V, Cr, Mn, Co, Ni, Cu, Zn, Y, Zr, and Ba). The wavelengths, species and excitation potentials of the atomic transitions employed in our study are reported in Table~\ref{linelist}. The last columns of the Table lists the EWs measured in the Solar spectrum by Casali et al. (submitted). This line list is based on the list employed in \citet{Melendez14}, that was assembled specifically for the analysis of solar twin stars by selecting preferentially unsaturated lines with minimal blending in the Solar spectrum. Equivalent widths of the atomic transitions listed in \citet{Melendez14} are measured with \texttt{Stellar diff}\footnote{\texttt{Stellar diff} is Python code publicly available at \url{https://github.com/andycasey/stellardiff}.}. This code allows the user to select one or more spectral windows for the continuum setting devoid of absorption features around each line of interest. We employ the same window settings to calculate continuum levels and fit the lines of interest with Gaussian profiles in all the exposures and the co-added spectrum of each star. The EW measurements are used by the qoyllur-quipu (q2) code \citep{Ramirez14} to determine the stellar parameters and chemical abundances for each exposure through the line-by-line differential analysis, using the co-added spectrum as a reference. The log R$^\prime_{\rm HK}$ indexes, EWs, atmospheric parameters and differential abundances determined for each single exposure from our analysis are listed in Tables~\ref{EWs}, \ref{params}, and \ref{abundances}. Note that the differential abundances reported in these tables are not relative to the Sun, but relative the co-added spectrum of the corresponding star.

\setcounter{table}{0}
\begin{table*}
\footnotesize
\centering
\caption{List of the 21,897 HARPS exposures employed in this study - full table available online at the CDS.}
\label{spectra}
\medskip
\begin{tabular}{ccccccc}
\hline
Spectrum ID & Star & Project ID & S/N  & Airmass & Exptime & BJD \\
& &  & [pxl$^{-1}$] & & [s] & \\
\hline
HARPS.2005-04-20T08:37:39.998 & $\alpha$ Cen A & 075.D-0800(A) & 397 & 1.40 & 3 & 2453480.86322119 \\
HARPS.2005-04-20T09:29:10.270 & $\alpha$ Cen A & 075.D-0800(A) & 258 & 1.56 & 3 & 2453480.89899796 \\
HARPS.2005-04-19T03:21:40.666 & $\alpha$ Cen A & 075.D-0800(A) & 370 & 1.28 & 5 & 2453479.64375465 \\
HARPS.2005-04-23T07:46:50.894 & $\alpha$ Cen A & 075.D-0800(A) & 339 & 1.31 & 2 & 2453483.82803507 \\
HARPS.2005-04-21T03:57:14.304 & $\alpha$ Cen A & 075.D-0800(A) & 385 & 1.22 & 4 & 2453481.66851639 \\
... & ... & ... & ... & ... & ... & ... \\
\hline
\end{tabular}
\end{table*}

\setcounter{table}{1}
\begin{table*}
\footnotesize
\centering
\caption{Line list - full table available online at the CDS.}
\label{linelist}
\medskip
\begin{tabular}{cccc}
\hline
Wavelength & Specie & $\chi_{\rm exc}$ & EW$_{\rm Sun}$ \\
 $[$$\AA$$]$ & & [eV] & $[$$m\AA$$]$\\
\hline
4365.896 & Fe~I & 2.990 & 51.1 \\
4445.471 & Fe~I & 0.087 & 40.5 \\
4602.001 & Fe~I & 1.608 & 71.7 \\
4779.439 & Fe~I & 3.415 & 40.5\\
4788.757 & Fe~I & 3.237 & 65.7\\
... & ... & ...  \\
\hline
\end{tabular}
\end{table*}

\setcounter{table}{2}
\begin{table*}[ht]
\footnotesize
\centering
\caption{Equivalent widths measured on each spectrum - full table available online at the CDS.}
\label{EWs}
\medskip
\begin{tabular}{cccccccccc}
\hline
Spectrum ID & Star  & log R$^\prime_{\rm HK}$ & $\lambda$4365.9 & err $\lambda$4365.9 & $\lambda$4445.5 & err $\lambda$4445.5 & $\lambda$4602.0 & err $\lambda$4602.0 & ... \\
& & [dex] & [m$\AA$] & [m$\AA$] & [m$\AA$] & [m$\AA$] & [m$\AA$] & [m$\AA$] & ... \\
\hline
HARPS.2005-04-20T08:37:39.998 & $\alpha$ Cen A & $-$5.154 & 59.18 & 0.17 & 49.14 & 0.20 & 81.03 & 0.17 & ...\\
HARPS.2005-04-20T09:29:10.270 & $\alpha$ Cen A & $-$5.160 & 59.31 & 0.17 & 49.29 & 0.22 & 80.12 & 0.17 & ...\\
HARPS.2005-04-19T03:21:40.666 & $\alpha$ Cen A & $-$5.146 & 59.20 & 0.20 & 48.96 & 0.22 & 79.66 & 0.16 & ...\\
HARPS.2005-04-23T07:46:50.894 & $\alpha$ Cen A & $-$5.151 & 59.70 & 0.20 & 49.94 & 0.24 & 80.57 & 0.20 & ...\\
HARPS.2005-04-21T03:57:14.304 & $\alpha$ Cen A & $-$5.148 & 59.37 & 0.20 & 49.09 & 0.20 & 80.48 & 0.17 & ...\\
... & ... & ... & ... & ... & ... & ... & ... & ... & ... \\
\hline
\end{tabular}
\end{table*}

\setcounter{table}{3}
\begin{table*}[ht]
\footnotesize
\centering
\caption{Atmospheric parameters determined for each exposure - full table available online at the CDS.}
\label{params}
\medskip
\begin{tabular}{ccccccccccc}
\hline
Spectrum ID & Star  & log R$^\prime_{\rm HK}$ & T$_{\rm eff}$ & err T$_{\rm eff}$ & log~g & err log~g & [Fe/H] & err [Fe/H] & $\xi$ & err $\xi$ \\
& & [dex] & [K] & [K] & [dex] & [dex] & [dex] & [dex] & [km s$^{-1}$] & [km s$^{-1}$]\\
\hline
HARPS.2005-04-20T08:37:39.998 & $\alpha$ Cen A & $-$5.154 & 5815 & 5 & 4.306 & 0.012 & 0.223 & 0.004 & 1.11 & 0.01\\
HARPS.2005-04-20T09:29:10.270 & $\alpha$ Cen A & $-$5.160 & 5817 & 5 & 4.316 & 0.013 & 0.222 & 0.004 & 1.11 & 0.01\\
HARPS.2005-04-19T03:21:40.666 & $\alpha$ Cen A & $-$5.146 & 5811 & 5 & 4.306 & 0.011 & 0.221 & 0.004 & 1.10 & 0.01\\
HARPS.2005-04-23T07:46:50.894 & $\alpha$ Cen A & $-$5.151 & 5808 & 5 & 4.286 & 0.012 & 0.228 & 0.004 & 1.08 & 0.01\\
HARPS.2005-04-21T03:57:14.304 & $\alpha$ Cen A & $-$5.148 & 5807 & 4 & 4.301 & 0.010 & 0.217 & 0.004 & 1.11 & 0.01\\
... & ... & ... & ... & ... & ... & ... & ... & ... & ... & ... \\
\hline
\end{tabular}
\end{table*}

\setcounter{table}{4}
\begin{table*}[ht]
\begin{threeparttable}
\centering
\footnotesize
\caption{Chemical abundances determined for each exposure  - full table available online at the CDS.}
\label{abundances}
\medskip
\begin{tabular}{cccccccc}
\hline
Spectrum ID & Star  & log R$^\prime_{\rm HK}$ & [C I/H] & err [C I/H] & [Na I/H] & err [Na I/H] & ... \\
& & [dex] & [dex] & [dex] & [dex] & [dex] &  ... \\
\hline
HARPS.2005-04-20T08:37:39.998 & $\alpha$ Cen A & $-$5.154 & 0.00 & 0.02 & 0.008 & 0.009 &  ... \\
HARPS.2005-04-20T09:29:10.270 & $\alpha$ Cen A & $-$5.160 & 0.015 & 0.007 & 0.02 & 0.02 &  ... \\
HARPS.2005-04-19T03:21:40.666 & $\alpha$ Cen A & $-$5.146 & 0.000 & 0.010 & 0.007 & 0.007 & ... \\
HARPS.2005-04-23T07:46:50.894 & $\alpha$ Cen A & $-$5.151 & $-$0.019 & 0.010 & $-$0.010 & 0.010 &  ... \\
HARPS.2005-04-21T03:57:14.304 & $\alpha$ Cen A & $-$5.148 & 0.06 & 0.06 & $-$0.002 & 0.011 &  ... \\
... & ... & ... & ... & ... & ... & ... & ...  \\
\hline
\end{tabular}
    \begin{tablenotes}
    \centering
      \footnotesize
      \item Note. The differential abundances reported in this table are relative to the co-added spectrum of the corresponding star.
    \end{tablenotes}
\end{threeparttable}
\end{table*}


\section{Results} \label{sec:Results}
In this Section we present our results and discuss how stellar activity affects the spectroscopic determination of atmospheric parameters and chemical abundances.

\subsection{Stellar activity and equivalent widths} \label{sec:lines}
\begin{figure}
\centering
\includegraphics[width=8cm]{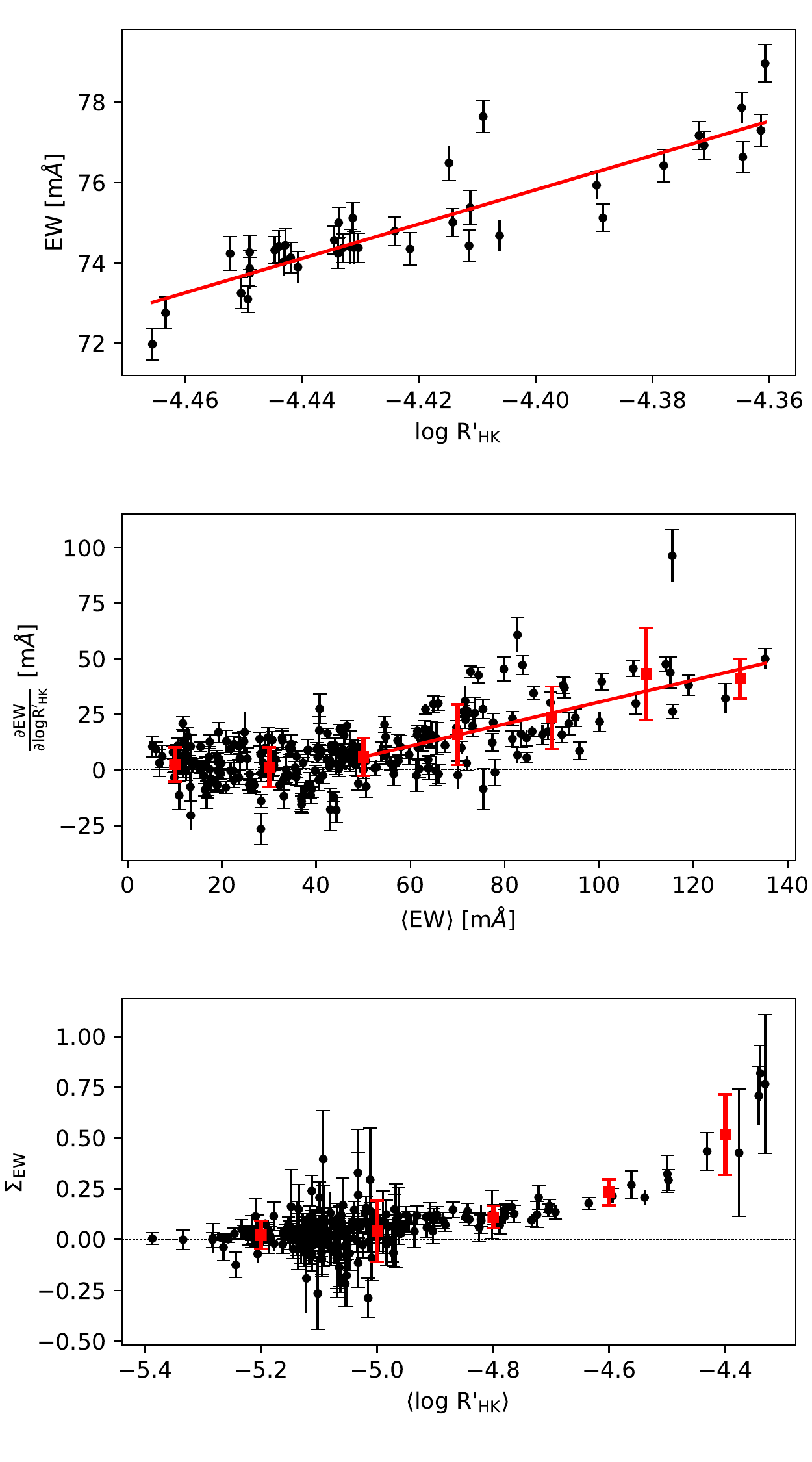}
\caption{\textbf{Top.} - Equivalent width of the Ba II line at 5853~$\rm\AA$ measured from the HARPS spectra acquired for HD59967 as a function of the stellar log R$^\prime_{\rm HK}$ index. The EW increases as a function of the stellar activity. The red solid line represents the linear fit of the distribution. \textbf{Middle.} - Black dots are the EW-log R$^\prime_{\rm HK}$ slopes (see top panel) of all the lines measured in HD59967 spectra as a function of the median line's EWs (i.e., $\langle$EW$\rangle$). The averaged values of these slopes at different bins of $\langle$EW$\rangle$ are plotted as red squares, while their error bars represent the standard deviations within the bin. Lines with $\langle$EW$\rangle$$\gtrsim$50~m$\rm\AA$ have EW-log R$^\prime_{\rm HK}$ slopes that increase with $\langle$EW$\rangle$. The linear fit of the distribution traced by absorption featured with $\langle$EW$\rangle$$>$50~m$\rm\AA$ is plotted as a red solid line. We refer to its slope value as $\Sigma_{\rm EW}$. \textbf{Bottom.} - Black dots are the $\Sigma_{\rm EW}$ for all stars in our sample as a function of their median log R$^\prime_{\rm HK}$ index. Red squares and their error bars represent the binned-averaged $\Sigma_{\rm EW}$ values the standard deviations at different intervals of $\langle$log R$^\prime_{\rm HK}\rangle$.
\label{EW_sensitivity}}
\end{figure}

The study by \citet{Galarza19} on the solar twin HD~59967 revealed that the EWs of iron lines can vary along the stellar activity cycle of quantities that depend on the mean line-centre optical depth $\tau_\lambda$. Namely, lines with log~$\tau_\lambda\geq$-1 do not show any significant variation along the cycle, but those that form in the external layers of the photosphere have EW values that change with chromospheric activity along the stellar cycle. This effect is also shown in Fig.~\ref{EW_sensitivity}-top where we plot the EW measurements of the Ba~II line 5853$\rm\AA$ measured from the HD~59967 spectra at different phases of its activity cycle: at higher activity levels, the EW is clearly higher. 

We perform a linear fit of the EWs-log R$^\prime_{\rm HK}$ relation for this and other atomic lines in the HD~59967 spectra. The resulting slopes are plotted in Fig.~\ref{EW_sensitivity}-middle as a function of $\langle$EW$\rangle$, the median EW of the line over all exposures. The $\langle$EW$\rangle$ value can be used as a proxy of the line optical depth $\tau_\lambda$, as stronger lines form higher above the optical surface of the stellar atmosphere \citep{Gray92}. The red squares in Fig.~\ref{EW_sensitivity}-middle panel and the relative error bars represent the binned averaged slope values and standard deviations calculated at different $\langle$EW$\rangle$ intervals. From the plot we observe that lines with $\langle$EW$\rangle\gtrsim$50~m$\rm\AA$ have slopes that are typically positive, indicating that their EWs increase with the activity index. All the other lines have slopes that are typically consistent with zero, meaning that their EWs have not varied significantly during the stellar cycle.


These trends are likely caused by stellar magnetic fields and cool spots affecting the formation of absorption lines observed in stellar spectra. Magnetic fields can affect spectral lines both directly, through the Zeeman effect, and indirectly, due to magnetically induced changes the temperature and density of the atmospheric plasma in the line formation region. In the spectrum of a star permeated by magnetic fields of some 10mT or more, the Zeeman pattern of many absorption lines is considerably wider than the thermal Doppler profile resulting in artificially larger EWs \citep{Babcock49}. The magnitude of magnetic intensification of the line depends on the strength of the magnetic field and on the $\tau_\lambda$ of the single line: absorption features that form near the top of the stellar photosphere, where magnetic fields are stronger, undergo a stronger magnetic intensification than lines that form in lower layers. Following the study of \citet{Babcock49} on magnetic intensification of stellar absorption lines, many theorists have suggested the presence of magnetic fields to explain the observed phenomenologies of pre-main-sequence stars, such as chemical anomalies or the lithium spread in young clusters (e.g., \citealt{Uchida84,Leone04,Leone07,Oksala18}). 

On the other hand, the indirect effects of magnetic fields arise from the fact that stellar photospheres become more transparent near magnetic concentrations due to the lower density of the plasma. This allows one to probe into deeper and hotter layers of the stellar atmosphere. The hotter temperatures that the radiation ``feels'' in these regions weakens the absorption lines with higher potential energy \citep{Fabbian12}. Since we do not observe weakening of lines as a function of the activity index, we consider that the magnetic intensification predicted by \citet{Babcock49} is a more likely explanation for the EW modulation than other indirect effects caused by magnetic fields. However, it is also possible that indirect effects due to the presence of strong magnetic fields in the vicinity of stellar spots and plage regions, have also affected the absorption lines creating a certain degree of scatter in the EW measurements, probably further modulated by the rotation of the star.

Finally, the same EW modulation can also be explained by the variation of the stellar surface covered by cold spots along the activity cycle. In fact, cold stellar spots can make the stellar photosphere appear cooler, increasing the EW of lines with low energy potential  \citep{Gray92}. Unfortunately, lines with low energy potentials tend to form at smaller $\tau_\lambda$, which means it is impossible to clearly determine if the main cause of the EW variation is the Zeeman broadening, cool stellar spots or a combination of the two. It is also possible that the relative importance of the direct and indirect effects of magnetic fields and stellar spots changes as the star ages, due to the drastic variation of spot filling factors, number of faculae, and strength of magnetic fields that stars undergo across the pre-main-sequence and early stages of the main sequence phases.

\begin{figure}
\centering
\includegraphics[width=8cm]{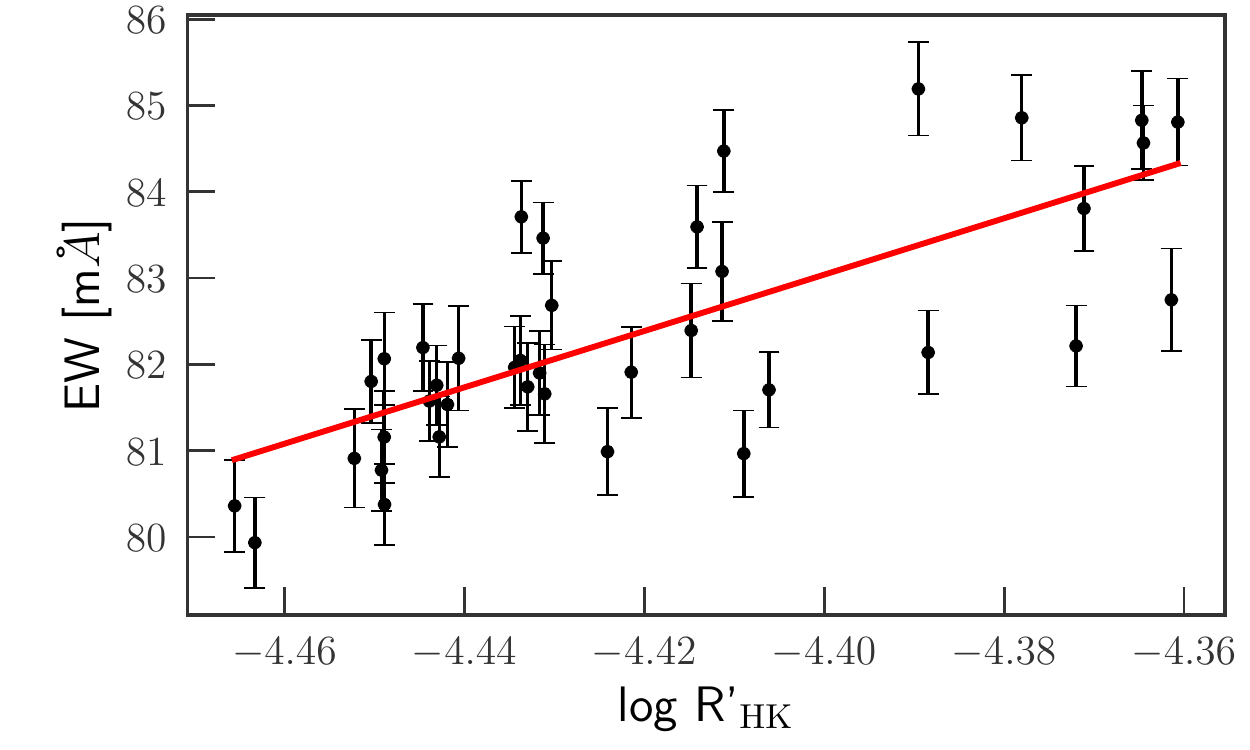}
\caption{Same of Fig.~\ref{EW_sensitivity}-Top, but here Voigt profiles have been used to measure the EWs of the Ba II line instead of Gaussians.}
\label{voigt}
\end{figure}

Our analysis confirms the conclusions given by \citet{Flores16} and \citet{Galarza19} for the solar twins HD59967 and HD45184, respectively. However, the goal of this paper is to pass from the analysis of single objects to the study of a larger number of stars of different ages and typical activity levels. To do so, we first perform a linear fit of the relation between the EW-log R$^\prime_{\rm HK}$ slopes and $\langle$EW$\rangle$ for all the absorption features with $\langle$EW$\rangle\geq$50~m$\rm\AA$. The resulting linear function is represented in Fig.~\ref{EW_sensitivity}-middle as a red solid line. The slope of this linear function is another important parameter that hereafter we call $\Sigma_{\rm EW}$. Similarly, we calculate $\Sigma_{\rm EW}$ for all the other stars in our sample. In Fig.~\ref{EW_sensitivity}-bottom we plot $\Sigma_{\rm EW}$ values along with their uncertainties as a function of $\langle$log R$^\prime_{\rm HK}$$\rangle$ for all stars in our sample, where $\langle$log R$^\prime_{\rm HK}$$\rangle$ is the median of the activity indexes measured at all epochs. The red squares represent the binned-averaged $\Sigma_{\rm EW}$ values and their standard deviations. The plot shows that the variation on EW during the activity cycle becomes significant for $\langle$log R$^\prime_{\rm HK}$$\rangle\gtrsim-$5.0 dex. Accordingly to the age-log R$^\prime_{\rm HK}$ relation calibrated by \citet{Lorenzo-Oliveira18} on solar twin stars, stars with log R$^\prime_{\rm HK}>-$5.0 dex are typically younger than $\sim$5 Gyr. The sensitivity of the EWs to the variation of chromospheric activity during the stellar cycle increases with the median stellar activity and, consequently, it is more significant for younger stars.

Another interesting outcome of this analysis is that the indirect effects of magnetic fields on stellar spectra are negligible in relation to the Zeeman broadening of atomic lines or the effect of cool stellar spots. This is a general behaviour of all the most active stars in our sample, regardless of other possible key factors, such as the morphology of magnetic fields, that can vary from star to star and that could determine the magnitude by which magnetic fields influence stellar spectra \citep{Moore15,Shchukina15}. 

Finally, we test whether Voigt profiles measure changes in EWs along the stellar cycle that are different than those obtained with Gaussian profiles. In Fig.~\ref{voigt} we show the EWs of the Ba II line at 5853~$\rm\AA$ measured with Voigt profiles in HD59967 spectra as a function of log R$^\prime_{\rm HK}$. These EWs are typically $\sim$8~m$\rm\AA$ larger than those obtained with Gaussians (i.e., see Fig~\ref{EW_sensitivity}-top), because a Voigt profile can better capture the damping wings of stronger lines. However, the EW variation as a function of log R$^\prime_{\rm HK}$ traced by Voigt profile is marginally consistent with that observed with Gaussians: while the first gave a EW-log R$^\prime_{\rm HK}$ slope equal to 33$\pm$5~m$\rm\AA$, the use of Gaussians produced a slope of 43$\pm$4~m$\rm\AA$.

\subsection{Stellar activity and stellar parameters} \label{sec:parameters}
In Section~\ref{sec:lines} we have shown that the modulation in chromospheric activity during the stellar cycle can modify the EW of absorption features of a multiplicative factor that depends on $\langle$EW$\rangle$. Stars with typical log R$^\prime_{\rm HK}>-$5.0 dex are affected by this phenomenon.

\begin{figure*}
\centering
\includegraphics[width=16cm]{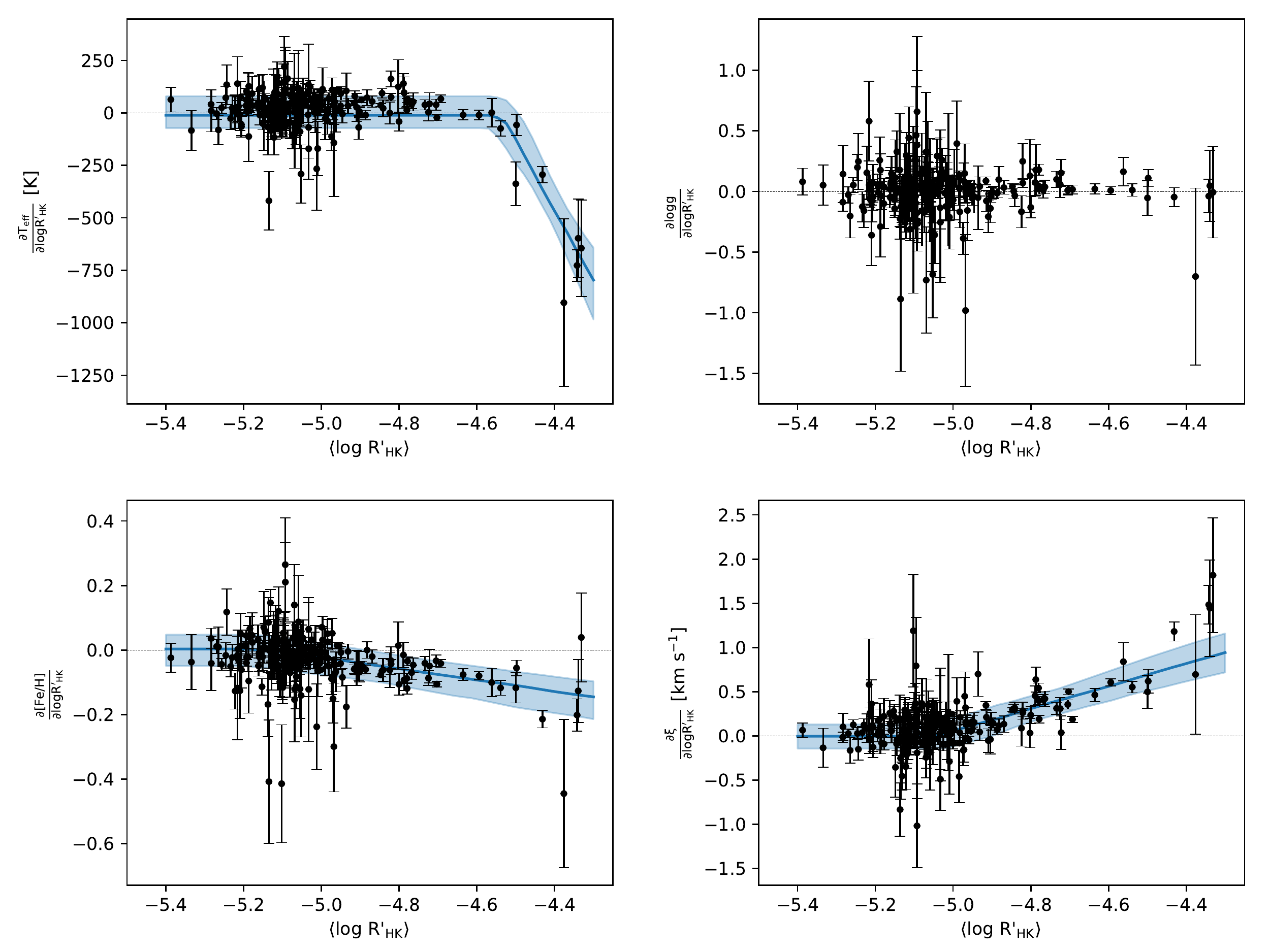}
\caption{The four panels show the sensitivity of the stellar parameters T$_{\rm eff}$, log~g, [Fe/H], and $\xi$ to the chromospheric activity for all the stars in our sample as a function of their typical log~R$^\prime_{\rm HK}$ index. Blue areas and the blue solid lines represent the posterior-probability 90$\%$ confidence intervals of MCMC simulation and the maximum posterior density estimates, respectively. 
\label{sensitivity_parameters_MCMC}}
\end{figure*}

As a consequence, the stellar parameters inferred from the simultaneous search for three spectroscopic equilibria of iron lines (i.e., excitation equilibrium, ionization balance, and the relation between log N$_{\rm Fe I}$ and the reduced equivalent width EW/$\lambda$) can also be indirectly influenced by the stellar chromospheric activity. This is clearly visible in the four panels of Fig.~\ref{sensitivity_parameters_MCMC}. Each panel shows the sensitivity of the four atmospheric parameters (T$_{\rm eff}$, log~g, [Fe/H], and $\xi$) to the variation in chromospheric activity. For example, $\partial$T$_{\rm eff}$/$\partial$log~R$^\prime_{\rm HK}$ as a function of $\langle$log~R$^\prime_{\rm HK}$$\rangle$ for all stars in our sample. A negative $\partial$T$_{\rm eff}$/$\partial$log~R$^\prime_{\rm HK}$ means that the T$_{\rm eff}$ value determined though our analysis for a particular star decreases as a function of chromospheric activity during the stellar cycle, while a $\partial$T$_{\rm eff}$/$\partial$log~R$^\prime_{\rm HK}$ consistent with zero indicates that the T$_{\rm eff}$ value has not changed during the stellar cycle.

As shown before, chromospheric activity can induce an increment of lines with typically large EWs. Therefore, higher $\xi$ values are required to balance the relation between abundances and EW/$\lambda$. The effect becomes more prominent at larger $\langle$log~R$^\prime_{\rm HK}$$\rangle$. This explains why $\partial\xi$/$\partial$log~R$^\prime_{\rm HK}$ increases with $\langle$log~R$^\prime_{\rm HK}$$\rangle$, as observed in the lower-right panel of Fig.~\ref{sensitivity_parameters_MCMC}. An increase of $\xi$ delays the saturation of the curve of growth of each absorption line and, as a consequence, decreases the inferred abundance of the corresponding element \citep{Gray92}. This effect is clearly visible in the lower-left panel of Fig.~\ref{sensitivity_parameters_MCMC}, where $\delta$[Fe/H]/$\delta$log~R$^\prime_{\rm HK}$ decreases with the increase of the stellar $\langle$log~R$^\prime_{\rm HK}$$\rangle$. The effect described in Section~\ref{sec:lines} also impacts the spectroscopic determination of  T$_{\rm eff}$ (see upper-left panel of Fig.~\ref{sensitivity_parameters_MCMC}), but it does not significantly affect log~g (see upper-right panel of Fig.~\ref{sensitivity_parameters_MCMC}).

A detailed inspection of Fig.~\ref{sensitivity_parameters_MCMC} can provide insights on how the chromospheric activity of stars affects our ability to infer T$_{\rm eff}$, [Fe/H], and $\xi$ from stellar spectra. With this aim, we simultaneously model the $\partial$P$_{i}$/$\partial$log~R$^\prime_{\rm HK}$ - $\langle$log~R$^\prime_{\rm HK}\rangle$ relations for the i$^{\rm th}$ atmospheric parameters P$_{i}$ through Markov-chain Monte Carlo simulations. For the procedure we adopt a model that switches between a null dependence from stellar activity at low log~R$^\prime_{\rm HK}$ values to a linear dependence $\partial$P$_{i}$/$\partial$log~R$^\prime_{\rm HK}$ from stellar activity at high log~R$^\prime_{\rm HK}$ values. The model is described as it follows:


\begin{equation}
\small
\frac{\partial P_{ij}}{\partial logR'_{\rm HK}} = \left\{
  \begin{array}{lr}
    \rm 0 & : \rm x_j < \tau_i\\
    \rm a_i\times(x_j -  \tau_i) & : \rm x_j \ge \tau_i
  \end{array}
\right.
\label{model_MCMC}
\end{equation}

where x$_{j}$ is the $\langle$log~R$^\prime_{\rm HK}$$\rangle$ value of the j$^{\rm th}$ star and $\tau_{i}$ is the switchpoint of the i$^{\rm th}$ parameter.
The model assumes priors for a$_{i}$ and $\tau$$_{i}$ that are Normal distributions $\mathcal{N}$($\mu$,$\sigma$), where $\mu$ is the mean and $\sigma$ the standard deviation. Namely, the priors for a$_{\rm T_{eff}}$, a$_{\rm [Fe/H]}$, and a$_{\rm \xi}$ are $\mathcal{N}$(5$\times$10$^{3}$ K, 10$\times$10$^{3}$ K), $\mathcal{N}$(0.0 dex, 3 dex), and $\mathcal{N}$(1 km s$^{-1}$, 5 km s$^{-1}$), respectively.  The prior for $\tau_{i}$ is  $\mathcal{N}$($-$5 dex, 1 dex).  We also assume that data points have Gaussian uncertainties of variance s$_{ij}^{2}$ + $\zeta_{i}^{2}$, where s$_{ij}$ is the uncertainty in $\partial$P$_{i}$/$\partial$log~R$^\prime_{\rm HK}$ for the j-star and $\zeta_{i}$ is a parameter that accounts for the possibility that s$_{ij}$ are underestimated and that HARPS observations have not homogeneously sampled the entire stellar cycles. The priors for $\zeta_{i}$ is a half Cauchy function with $\gamma$ parameter equal to 10 K, 1 dex, and 10 km s$^{-1}$ for T$_{eff}$, [Fe/H] and $\xi$ respectively. We ran the simulation with 10,000 sample, half of which are used for burn-in, and employing the No-U-Turn Sampler \citep{Hoffman11}. The script was written in Python using the {\tt pymc3} package \citep{Salvatier16}.

The convergence of the simulations have been checked by inspecting the traces for each parameter and their autocorrelation plots. The 90$\%$ confidence intervals of the posteriors are listed in Table~\ref{posteriors}: they are well within the ranges allowed by the priors. The 90$\%$ confidence intervals of the models resulting from the inference are represented in Fig.~\ref{sensitivity_parameters_MCMC} as blue areas. 

\setcounter{table}{5}
\begin{table*}
\centering
\caption{Model posteriors.}
\label{posteriors}
\medskip
\begin{tabular}{cccc}
\hline
Parameter & a &$\tau$ & $\zeta$ \\
 & [5$\%$, 50$\%$, 95$\%$] & [5$\%$, 50$\%$, 95$\%$] & [5$\%$, 50$\%$, 95$\%$] \\ \hline
T$_{\rm eff}$ [K] & $-$5056, $-$3510, $-$2109 & $-$4.58, $-$4.53, $-$4.48 & 40.4, 47.5, 55.0 \\
$[$Fe/H$]$ [dex]& $-$0.221, $-$0.172, $-$0.123 & $-$5.23, $-$5.15, $-$5.07 & 0.024, 0.029, 0.034 \\
$\xi$ [km s$^{-1}$] & 1.02, 1.30, 1.61 & $-$5.10, $-$5.05, $-$4.98 & 0.082, 0.100, 0.118 \\
\hline
\end{tabular}
\end{table*}

In our model, the spectroscopic determination of the P$_{i}$ parameter is influenced by activity cycle only if the star has a  $\langle$log~R$^\prime_{\rm HK}\rangle$ $\geq$ $\tau_{i}$. The $\tau$ switchpoint inferred for T$_{\rm eff}$ occurs with the highest probability at $\sim$$-$4.5 dex, which corresponds to solar twins of 1 Gyr \citep{Lorenzo-Oliveira18}. While the switchpoint inferred for [Fe/H] and $\xi$ happens at $\sim$$-$5.0, meaning that the determination of these parameters of solar twins younger than 4-5 Gyr is possibly affected by magnetic fields or stellar spots. However, it must be noted that these results are heavily dependent on the master list of absorption features employed in the spectroscopic analysis. In fact, a reduced use of absorption lines with $\langle$EW$\rangle$$\leq$50~m$\rm\AA$ in the spectroscopic analysis would mitigate the impact of magnetic fields or spots on the determination of stellar parameters. However, in this case it would be more difficult to probe microturbulence due to a lack of strong lines.

Finally, we can predict how a star identical to the Sun (i.e., T$_{\rm eff}$=5770 K, [Fe/H]=0.0 dex, $\xi$=1.0 km s$^{-1}$) would look as a function of its chromospheric activity if analysed with the same linelist employed in this study. 
To do so, we assume that the variation of the i$^{\rm th}$ parameter for the  j$^{\rm th}$ star $\Delta$P$_{ij}$ is equal to zero if the stellar activity index is smaller than the switchpoint (i.e., x$_{j}$$<$$\tau_{i}$), while for larger activity indexes it is equal to the integral of Eq.~\ref{model_MCMC}. Our model also includes a second term that depends on $\zeta_{i}$ and that accounts for the possibility that the measured uncertainty $\partial$P$_{i}$/$\partial$log~R$^\prime_{\rm HK}$ is underestimated. Therefore:

\begin{equation}
\small
\Delta P_{ij} = \left\{
  \begin{array}{lr}
    \rm 0 & : x_j < \tau_i\\
    \rm \frac{1}{2}a_i \times (x_j -  \tau_i)^2 + \zeta_i \times (x_j -  \tau_i) & : x_j  \ge \tau_i
  \end{array}
\right.
\label{prediction}
\end{equation}

\begin{figure*}
\centering
\includegraphics[width=16cm]{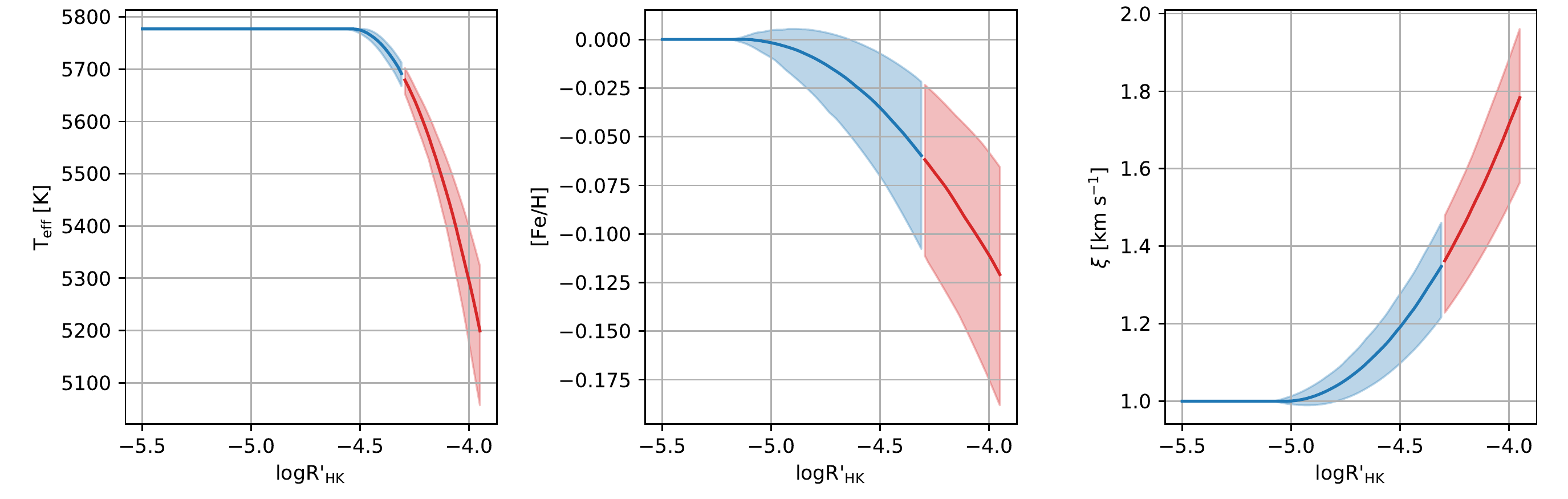}
\caption{The three panels describe how the stellar parameters T$_{\rm eff}$, [Fe/H], and $\xi$ vary as a function of the stellar activity based on the model described in Eq.~\ref{prediction} and the posteriors' distributions obtained through the MCMC simulation. The blue areas represent the 90$\%$ confidence intervals within the rage of log~R$^\prime_{\rm HK}$ indexes covered by the stars in our sample, while the red areas are the extrapolations of the same model at higher log~R$^\prime_{\rm HK}$ values. The solid lines represent the maximum a posteriori estimates. No variation is expected for log~g.
\label{MCMC_prediction}}
\end{figure*}

The $\Delta$P$_{i}$ probability distribution at each log~R$^\prime_{\rm HK}$ has been inferred from the a$_{i}$, $\tau_{i}$, and $\zeta_{i}$ posteriors resulting from the MCMC simulation. In Fig.~\ref{MCMC_prediction} we show the 90$\%$ confidence intervals of these distributions. The blue areas highlight the range in log~R$^\prime_{\rm HK}$ values covered by our dataset, while the red areas are an extrapolation at larger activity indexes of the model fitted to the observations. The maximum variations in stellar parameters traced by blue areas are around 100~K, 0.06 dex and 0.35 km~s$^{-1}$ for T$_{\rm eff}$, [Fe/H], and $\xi$, respectively. These values are similar to the typical uncertainties in atmospheric parameters provided by large spectroscopic surveys observing at optical wavelengths, such as the Gaia-ESO and GALAH surveys \citep{Smiljanic14,Buder2018}. However, these surveys targeted pre-main-sequence clusters and stars that are significantly more active than those analysed here. Thus, even if spectroscopic surveys use different methods of analysis and linelists than those employed in our study, it is possible that the magnetic fields or star spots have affected to a certain extent some scientific outcomes of these collaborations (see Section~\ref{sec:puzzles} for a discussion on the scientific implications of our result). 

Since our dataset samples the artificial variation of stellar parameters up to activity indexes of log~R$^\prime_{\rm HK}$$\sim$$-$4.3 dex, the red areas in Fig.~\ref{MCMC_prediction} represent an extrapolation of this phenomenon at higher indexes, up to log~R$^\prime_{\rm HK}$$\sim$$-$4.0 dex, values that are typically measured in star forming regions \citep{Mamajek08}. Interestingly, $\xi$ values measured for members of these young associations are typically around 2 km s$^{-1}$ (e.g., \citealt{James06,Santos08}), in agreement with the prediction in Fig.~\ref{MCMC_prediction} - right panel. This indicates that magnetic fields and stellar spots play an important role in shaping atomic lines in stellar spectra also at the high activity regimes typical of pre-main-sequence stars. Also the prediction of a T$_{\rm eff}$ variation of $\sim$600 K at log~R$^\prime_{\rm HK}$$\sim$$-$4.0 dex is partially consistent with the results of other studies that compared spectroscopic and photometric T$_{\rm eff}$ values for stars in pre-main-sequence stars \citep{Morel03,Baratella20} finding differences up to 400~K. However, these studies are not able to probe the effect of stellar spots, which have a roughly similar quantitative effect on the temperatures derived from spectroscopic and photometric data (see \citealt{Fekel86,Morel03}). Thus, the question around the full effect of stellar activity on spectroscopic T$_{\rm eff}$ of T Tauri stars is still open. For instance, it is possible that in extremely active stars (e.g., log~R$^\prime_{\rm HK}\sim$-4.0 dex) the Zeeman effect and the consequences of cold stellar spots are counteracted but not fully compensated by the indirect effects that magnetic fields have on the stellar photosphere, which allow stellar spectra to probe into deeper layers where temperatures are higher. This could be possible due to the significantly higher strength of magnetic fields on T Tauri stars compared to other objects older than a few 10 Myr.

\subsection{Stellar activity and chemical abundances} \label{sec:abundances}
The selective intensification of the strongest absorption lines in stellar spectra affects the determination of chemical abundances in two opposite ways. On one side, the EW intensification produces an increase of chemical abundances. On the other hand, the growth of $\xi$ has the effect of lowering the chemical abundances. Therefore, the net effect on abundances depends on the relative importance of these two reactions to stellar activity, which in turn depends on the linelist that is used for the analysis. 

\begin{figure*}
\centering
\includegraphics[width=16cm]{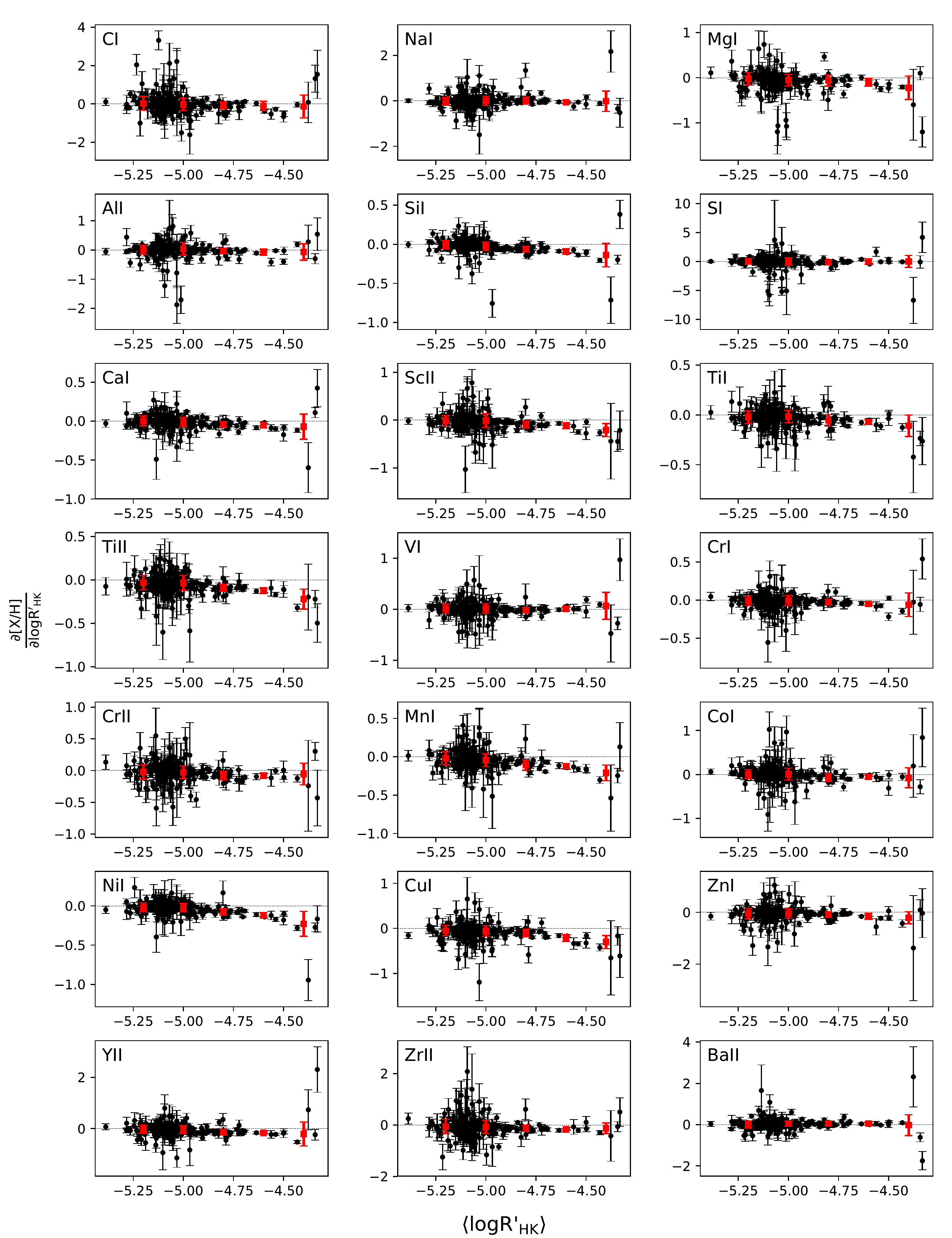}
\caption{Each panel shows the sensitivity of the X-element abundance to chromospheric activity (i.e., $\partial$[X/H]/$\partial$log~R$^\prime_{\rm HK}$) for all stars in our sample as a function of their typical log~R$^\prime_{\rm HK}$ index (black dots). Red squares and their error bars represent the binned-averaged $\partial$[X/H]/$\partial$log~R$^\prime_{\rm HK}$ values and standard deviations at different intervals in $\langle$log~R$^\prime_{\rm HK}\rangle$.
\label{elements}}
\end{figure*}

In Fig.~\ref{elements} we show the dependence of elemental abundances from stellar activity $\partial$[X/H]/$\partial$log~R$^\prime_{\rm HK}$ as a function of $\langle$log~R$^\prime_{\rm HK}\rangle$. While each black dot represents one star of the sample, the red squares and their error bars correspond to the binned averaged $\partial$[X/H]/$\partial$log~R$^\prime_{\rm HK}$ value and its standard deviation at different $\langle$log~R$^\prime_{\rm HK}\rangle$ intervals. These values are weighted by the $\partial$[X/H]/$\partial$log~R$^\prime_{\rm HK}$ uncertainties. The general behaviour traced by the red symbols shows that the growth in $\xi$ compensates the abundance increase for all the elements. Similarly to what we have observed for [Fe/H] (see Fig.~\ref{sensitivity_parameters_MCMC}), the variation of $\xi$ along the cycle of the most active stars is large enough to lower the abundances of Si, Sc, Ti, Mn, Ni, and Cu. This is not surprising, in fact, according to our linelist, these elements are mostly determined through the measurement of absorption lines that form deep into the stellar photosphere (i.e., $\tau_\lambda$$>$1) and that are not significantly intensified by the stellar magnetic fields or cold spots. Furthermore, most of these lines are medium-strong (e.g., EW$\sim$40-70~m$\rm \AA$) and sensitive to $\xi$. Therefore, for these elements the increase of line EWs is not enough to compensate for the growth in $\xi$.

There is a second class of elements formed by C, Na, Al, S, V, Co, Y, and Zr for which the lines used are also formed deep in the atmosphere, but these lines are so weak that they are not sensitive to $\xi$. Therefore, the change in the abundances of these elements as a function of $\langle$log~R$^\prime_{\rm HK}\rangle$ is nearly zero.

Finally, we identify a third group of elements (i.e., Mg, Ca, Cr, and Ba) that are detected mostly through strong lines formed in the upper atmosphere. These lines are both sensitive to $\xi$ and intensified by the stellar magnetic fields or cold spots, in a way that one effect counterbalance the other. Therefore, their change in abundances is also nearly zero.

\section{Scientific implications} \label{sec:puzzles}
In the previous section we have shown how chromospheric activity can affect the stellar parameters and elemental abundances derived from stellar spectra due to the magnetic broadening of the absorption lines and cold stellar spots with $\langle$EW$\rangle$$\gtrsim$50~m$\rm\AA$. This result provides a definitive explanation to important open questions in the study of the chemical evolution of the Galaxy (e.g., the low metallicity of the local ISM, the metal content of Orion, and the Ba puzzle) as well as key implications for chemical tagging, planet hunting, and many other studies based on stellar parameters determined from spectra of young stars. Below we discuss the significance of our result to these topics.

\subsection{The anaemia of the local ISM}
The chemical analysis of young ($\lesssim$100~Myr) stars is extremely important in the context of Galactic chemical evolution as it provides strong constraints to the models of stellar nucleosynthesis. In addition, young stars have not had time to disperse along the Galactic disk, therefore their chemical content is representative of the ISM's composition at the location where they are observed today. In contrast to models of Galactic chemical evolution (e.g., \citealt{Minchev13,Sanders15,Frankel18}), independent studies have consistently found that the youngest stars in our Galaxy have a metal content lower than the Sun. For instance, the metal content measured in different star forming regions located within 500 pc from the Sun (i.e., Chamaeleon, Corona Australis, Lupus, Orion Nebula Cluster, Rho Ophiuchi, Taurus) is on average equal to [Fe/H]=$-$0.07$\pm$0.03 dex \citep{Cunha98,Santos08,DOrazi11,Biazzo11a,Biazzo12a,Spina14,Spina17}, which is $\sim$15$\%$ lower than the Solar metallicity. Interestingly, members of star forming regions are found to have $\xi$ values significantly higher than that of the Sun and typically within the range 2.0-2.5 km s$^{-1}$ (e.g., \citealt{James06,Santos08,Baratella20}). The typical log~R$^\prime_{\rm HK}$ index of these stars is $\sim$$-$4.0 dex \citep{Mamajek08b}. The comparison of these values to the prediction in Fig.~\ref{MCMC_prediction} suggests that the youngest stars of the solar vicinity appear to be artificially metal poor and with high $\xi$ as a consequence of their high activity levels (e.g., strong magnetic fields, high coverage of cold stellar spots) that have selectively intensified the strongest atomic lines. According to our predictions, these star forming regions should actually contain the same amount of metals of the Sun.

\subsection{The metal content of Orion}
The [Fe/H]-log~R$^\prime_{\rm HK}$ prediction in Fig.~\ref{MCMC_prediction} also provides a convincing explanation for the metal content measured in the sub-clusters of the Orion association. Orion is one of the nearest regions (d$\sim$350-450 pc) of ongoing star formation where both low and high mass stars are formed. It is a complex composed of different sub-clusters with different ages, as the stellar formation burst has spread across the association during the last 20 Myr, triggered by supernovae explosions \citep{Bally08}. Since type II supernovae are sites of major nucleosynthesis, these explosions may also chemically enrich parts of the surrounding interstellar gas, and hence the newly formed next generation of stars (e.g., \citealt{Reeves72,Cunha92,Cunha94}). Therefore, one would expect that the sequential star formation occurring in Orion should result in a peculiar chemical enrichment with the youngest regions being enhanced in metals relative to older ones. Instead, the Orion Nebula Cluster, which is the youngest region of the Orion association, has been found to be the metal-poorest \citep{DOrazi09b,Biazzo11a,Biazzo11b}. Our analysis suggests a revised analysis of Orion's metal content that should properly take into account the effects of magnetic fields and cold spots on stellar spectra. Providing the first evidence of self-enrichment in a young stellar association would give fundamental insights into stellar nucleosynthesis and, most importantly, on the role that supernovae explosions have in the sequential collapse of molecular clouds, hence on the origin of stars and stellar clusters.

\subsection{The barium puzzle}
The so-called $barium$ $puzzle$ is still one of the most debated open questions around the production of $s$-process elements in the Milky Way. It originated when \citet{DOrazi09} measured [Ba/Fe] ratios in young ($<$50 Myr) open clusters in the solar vicinity ($<$500 pc) which showed to have $\sim$0.3 dex higher Ba than the value predicted by models of stellar nucleosynthesis \citep{Travaglio99,Busso01}. Further, in contrast to the anomalous Ba overabundance, the abundances of other $s$-process elements such as Y, Zr, La and Ce relative to Fe was found to be Solar \citep{DOrazi12}. Interestingly, further studies in young open clusters have shown that additional channels of nucleosynthesis, such as the intermediate neutron-capture process \citep{Cowan77}, cannot explain the Ba overabundance compared to other neutron-capture elements \citep{Mishenina15}.

A new piece of this puzzle was provided by \citet{Reddy15}, who analysed five young (5-200 Myr) local associations and found that they cover an abnormally large range [Ba/Fe] ratios, from +0.07 to +0.32 dex. A further analysis of solar twin stars by \citet{Reddy17} finally provided some clues to the solution of the $Ba$ $puzzle$. Namely, they showed a trend of increasing abundances from the Ba II 5853 $\rm\AA$ line with stellar activity among coeval stars. Therefore, they speculated that the high Ba abundance measured in young associations is not nucleosynthetic in origin but associated with the level of stellar activity. Specifically, they argued that a $\xi$ value derived from Fe lines that form at much larger $\tau_\lambda$ in the photosphere is not sufficient to represent the true broadening imposed by the turbulence of the upper photospheric layers where the Ba II lines form.

Our analysis proceeds in this framework, demonstrating that a relation exists between the EW of lines  formed at small $\tau_\lambda$ into the stellar atmosphere, such as the Ba II lines, and the stellar activity (see Fig.~\ref{EW_sensitivity}, top and middle panels). This dependence is visible in our data only for stars with  log R$^\prime_{\rm HK}\gtrsim$-5.0 dex (Fig.~\ref{EW_sensitivity}, bottom panel). 

However, in apparent contradiction to Fig.~\ref{EW_sensitivity}, Fig.~\ref{elements} does not show any clear evidence of a systematic positive dependence of Ba abundances from stellar activity. This is not surprising because, as we pointed out in Section~\ref{sec:abundances}, the increase in $\xi$ as a function of stellar activity has the important consequence of lowering chemical abundances that are derived at high activity levels. Therefore, in our analysis, the increase in $\xi$ is large enough to counterbalance the effect that the Zeeman broadening and stellar spots would have on Ba abundances. 

Since the increment in $\xi$ is highly dependent on the line list employed in the analysis and in particular to the number of lines with high EW/$\lambda$ that can trace the turbulence in the upper stellar layers, the use of different line lists can result in a different sensitivity of the $\xi$ parameter to stellar activity. Therefore, a different line list can also produce very different abundances of Ba, whose lines are extremely sensitive to $\xi$. This explains the large spread in Ba abundances found in young nearby associations by different teams \citep{DOrazi09,Reddy15,Reddy17}. For instance, the $\xi$ applied by \citet{Reddy17} for the calculation of Ba abundances was taken from \citet{Nissen15} and based on a list of weak Fe (EWs$\leq$70~m$\rm \AA$) formed quite deep in the atmosphere. Therefore, they were using $\xi$ values that do not reflect the extra broadening of absorption lines in the upper layers of active stars, where Ba lines are formed. On the other hand, our line list includes Fe lines with formation depths similar to those of the Ba lines. This explains the apparent contradiction of the large Ba abundances obtained by \citet{Reddy17} with the lack of Ba variation in Fig.~\ref{elements}. In a similar way, while our masterlist contains 78 Fe I lines, the one used by \citet{DOrazi09} was probably too small (only 33 Fe I lines) to adequately probe the turbulence in the upper stellar layers. In fact, while KG-type stars younger than 50 Myr have $\xi$ that are typically greater than 1.5 km s$^{-1}$ (see also \citealt{DOrazi12}), the $\xi$ values estimated in \citet{DOrazi09} are within 0.7 and 1.2 km s$^{-1}$ and very close to their first guess values. In conclusion, the Zeeman effect or stellar spots have intensified the lines used to determined the Ba abundance by \citet{DOrazi09}, but in their analysis this effect was not counteracted by any $\xi$ increase as in our analysis, leaving the Ba abundances anomalously high.

\subsection{Chemical signatures of planet engulfment events}
Do stars swallow their own planets? The major consequence of planet engulfment would be a chemical enhancement of the host star due to the pollution of rocky material. If the accreting star has a sufficiently thin convective zone, the planetary material is not too diluted and can produce a significant increase of the atmospheric metallicity, which can be reliably detected \citep{Spina15,Church19}. In fact, such dilution will not yield an indiscriminate abundance rise of all the metals, but likely will produce a characteristic chemical pattern that mirrors the composition observed in rocky objects with mostly refractory elements  (i.e., those with higher condensation temperature) being overabundant relatively to volatiles \citep{Chambers10}.

The chemical signatures of planet engulfment events have been found among members of binary systems (e.g., \citealt{Ramirez15,Teske16,Oh18,TucciMaia18,Nagar19}) and open clusters \citep{Spina15,Spina18,DOrazi19}. Members of the same stellar association are born at the same time and from the same gas, therefore they should be chemically identical. This indicates that the chemical anomalies found among members of the same association cannot be explained by processes of nucleosynthesis. However, this work poses the suspicion that these chemical anomalies are not actually due to planet engulfment events, but instead caused by different activity levels of the members of the same stellar association. 

\begin{figure}
\centering
\includegraphics[width=8cm]{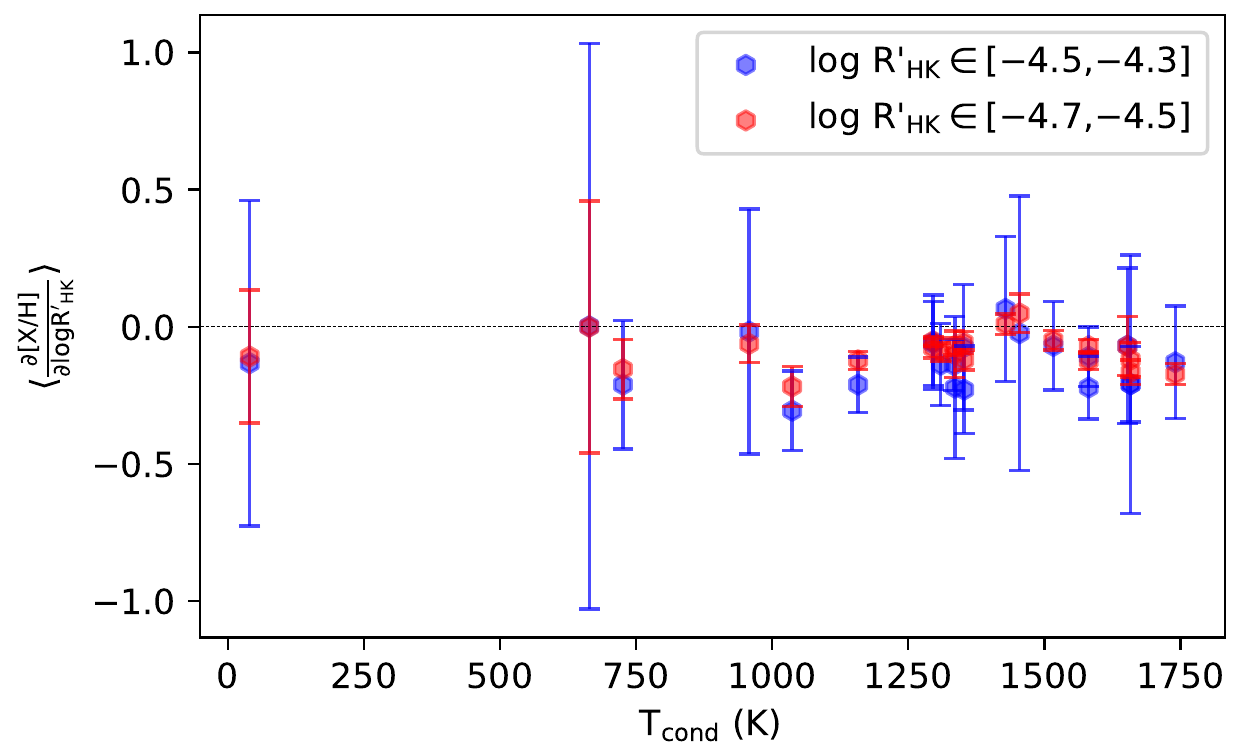}
\caption{The plot shows the binned-averaged $\partial$[X/H]/$\partial$log~R$^\prime_{\rm HK}$ values and standard deviations calculated for the intervals log~R$^\prime_{\rm HK}\in$[$-$4.7,$-$4.5] (red) and log~R$^\prime_{\rm HK}\in$[$-$4.5,$-$4.3] (blue) as a function of the condensation temperature T$_{\rm cond}$ of the X-element.
\label{Tcond_sensitivity}}
\end{figure}

The red and blue symbols in Fig.~\ref{Tcond_sensitivity} represent the average of the $\partial$[X/H]/$\partial$log~R$^\prime_{\rm HK}$ values for stars with log~R$^\prime_{\rm HK}\in$[-4.7,-4.5] and log~R$^\prime_{\rm HK}\in$[-4.5,-4.3], respectively, as a function of the condensation temperature T$_{\rm cond}$ of the X-element listed in \citet{Lodders03}. From this plot it is evident that some elements are sensitive to the variation in chromospheric activity during the stellar cycle, while others are not. We also observe that there is no clear relation between the sensitivity of an element to the stellar activity and its condensation temperature. Therefore, from these data there is no evidence in support of the possibility that chromospheric activity could mimic signatures of planet engulfment events in the chemical composition of stars. Instead, there are elements suggesting that the chemical anomalies found so far are genuine and not related to stellar activity. For example, a number of chemically anomalous Sun-like stars are older than 5 Gyr \citep{Ramirez15,Ramirez19,TucciMaia19}, therefore - according to our results - their activity levels are not high enough to produce the observed differences in elemental abundances, i,e., $\Delta$[Fe/H]$\geq$0.05 dex. Furthermore, even when the anomaly has been found among members of young stellar associations, there is no relation between the Fe abundance of the stars and their $\xi$ value \citep{Spina15,Spina18,DOrazi19}, as one would expect if the chemical anomaly was driven by an especially high (or low) activity level of the anomalous star compared to the other siblings.

\subsection{Chemical tagging}

Similar to a DNA profile, one could use the individual chemical patterns of stars that are not in clusters today to trace them back to a common site of origin \citep{Freeman02}. This approach - commonly called ``chemical tagging'' - is a powerful tool for Galactic archaeology, which aims at recovering the remnants of the ancient building blocks of the Milky Way (e.g., clusters, super-clusters, moving groups) that are now dispersed, reconstructing their star formation history and the migration rate of stars within the Milky Way (e.g., \citealt{BlandHawthorn10}). In fact, the main motivation behind the large-scale spectroscopic surveys of the current decade (e.g., APOGEE, Gaia-ESO, GALAH; \citealt{Holtzman15,Gilmore12,DeSilva15}) has been the acquisition of large and homogeneous sets of spectroscopic data from different environments within the Galaxy to trace its history in space and time.

Regardless of the precision achievable in elemental abundances, the success of chemical tagging relies on the significance of critical factors, including the level of chemical homogeneity within members of open clusters and the chemical diversity between open clusters. Even if it is now established that processes of atomic diffusion \citep{Dotter17} and planet engulfment events \citep{Laughlin97} can imprint chemical inhomogeneities among members of the same stellar association, a growing number of studies based on high-precision analysis of solar twin stars are showing that most cluster members on the same evolutionary phase are chemically identical at the typical precision levels reached by large spectroscopic surveys \citep{Liu16a,Liu16b,Spina18b,Nagar19}. 

On the other hand, our analysis shows that magnetic fields or stellar spots can reduce the chemical diversity between stars of different ages, posing a serious challenge for chemical tagging. In fact, while younger stars in our Galaxy should be chemically richer, they also tend to  appear poorer in metals than older stars due to the increasing levels of stellar activity (see Fig.~\ref{MCMC_prediction} - middle panel). In fact, recent studies have shown how challenging is to reconstruct and reassemble the dissolved stellar associations in the Solar vicinity or identify the dispersed family of open clusters solely based on the chemical composition of stars (e.g., \citealt{BlancoCuaresma18,Ness18,Simpson19,Casey19}). Therefore, new methods of spectroscopic analysis that could consider the effect of chromospheric activity in the stellar spectra (e.g., \citealt{Baratella20}) would pave the way for chemical tagging in the Milky Way disk.

On the other hand, in the context of chemical tagging, it is also possible to use the effect line intensification on stellar spectra to our advantage. In fact, the evidence that equivalent widths of lines can vary during the stellar cycle of quantities which depend on $\tau_\lambda$, allows us to use abundance ratios from lines of the same element formed at different depths in the stellar atmosphere to identify young and active stars in the field or among candidate members of young associations.

\subsection{Stellar ages}
It is well known that a star, as it ages, evolves along a determined track in the Hertzsprung-Russell diagram that - to a first approximation - depends on the stellar mass and metallicity. Therefore, if the atmospheric parameters and the absolute magnitude M$_{\rm V}$ of the star are known with enough precision, it will be possible to determine reasonable estimates of its age and mass \citep{Vandenberg85,Lachaume99}. In Fig.~\ref{MCMC_prediction} we show that Sun-like stars with higher chromospheric activity tend to appear cooler and metal poorer than what they really are. As a consequence, their ages and masses determined though isochrones are systematically overestimated (see also \citealt{Galarza19}). This has fundamental implications for all the studies that aim to trace the nucleosynthetic history of elements in the Galaxy through stellar [X/Fe]-age relations and chemical clocks (e.g., \citealt{Nissen15,Nissen16,Spina16,Spina16b,Spina18,Feltzing17,TucciMaia16,Bedell18}).

\subsection{Interstellar extinction}
In the recent years, spectroscopic and photometric Galactic surveys have enabled the computation of three-dimensional interstellar extinction maps thanks to accurate stellar atmospheric parameters and line-of-sight distances. This can be achieved by comparing the observed colours to those computed through the stellar parameters and a set of isochrones (e.g., \citealt{Schultheis15,Schlafly17}). This technique is extensively used also to infer interstellar extinction, which is particular important for young stars in star forming regions or pre-main-sequence clusters that are still partially embedded in the parental cloud. In fact, knowledge of the bolometric luminosities of stars in these young associations is vital for a broad variety of topics in stellar astrophysics. These include the studies of mass segregation, dynamical and structural properties of stellar associations before their dissolution, age spread in star forming regions, rate of star formation in giant molecular clouds and its dependence on time and stellar mass, the analysis of photoevaporation of circumstellar disks and its impact on planet formation (e.g., \citealt{Luhman08,Sacco17,Prisinzano19}).

Young stars are very active and their spectra are heavily affected by the strong magnetic fields. According to our analysis, the T$_{\rm eff}$ of these stars is systematically underestimated due to the effect of chromospheric activity (Fig.~\ref{MCMC_prediction} - left panel), which results in an underestimation of the stellar extinction, hence to an overestimation of the bolometric luminosity. Therefore, depending on the technique of spectroscopic analysis and the employed line list, stellar activity may have affected the results of observational studies of pre-main-sequence populations.

\subsection{Planet hunting}
The possibility to disentangle the effect of stellar activity and jitter from the radial velocity modulation of stars is of paramount importance for spectroscopic surveys that aim at planet detection. This can be achieved through different indicators, such as the bisectors of the spectral cross-correlation function \citep{Queloz01}, H$\alpha$ \citep{Bonfils07,Robertson14} and log~R$^\prime_{\rm HK}$ \citep{Noyes84,Delisle18}. More recently, new approaches using these activity indicators and statistical techniques such as Gaussian Processes (e.g., \citealt{Haywood14,Rajpaul15,Jones17,Delisle18}) or Moving Average (e.g., \citealt{Tuomi13}) have significantly improved our ability to mitigate the impact of stellar activity on the planetary signal. Recent works have searched for new activity indicators, showing how stellar activity affects spectral lines in different ways \citep{Thompson17,Davis17,Wise18,Zhao20}, which opens up the possibility of using the wealth of information contained in high-resolution spectra to verify the authenticity of a planetary signal (e.g., \citealt{Dumusque18}).

Our analysis has shown that not all lines have the same sensitivity to stellar activity, but that only the strongest ones (i.e., those with EW$>$50 m$\rm\AA$ or with log~$\tau_\lambda$$<$-1~dex; see also \citealt{Galarza19}) vary their EW during the activity cycle (see Fig.~\ref{EW_sensitivity}). Therefore, by masking all the lines for which no variation is expected along the stellar cycle and by detecting all the other lines, it is possible to further reduce the noise from stellar activity. This is a very promising possibility that could significantly increase the potential of high-resolution spectrographs in the hunt for planets around Sun-like stars. This is especially true for the spectrographs that do not observe the Ca lines, such as Veloce-Rosso \citep{Gilbert18} and Minerva-Australis \citep{Addison19}.

\section{Summary} \label{sec:conclusions}
In this study we analyse 21,897 HARPS spectra of 211 Sun-like stars. These stars are observed at high-resolution (R$\sim$115,000) and high S/N ($\geq$100 pixel$^{-1}$) at different phases of their activity cycles. The main goal of this experiment is to provide a quantitative evaluation of the effect that chromospheric activity has on the atmospheric parameters and elemental abundances that we infer from stellar spectra. Our main results can be summarised as follows:
\begin{itemize}
\item The EWs of spectroscopic lines increase with chromospheric activity along the stellar cycle of quantities that depend on the median EW of the line (Fig.~\ref{EW_sensitivity}-top). Specifically, we observe the largest variation for lines with the highest EW (Fig.~\ref{EW_sensitivity}-middle).

\item This effect is visible for stars with log R$^\prime_{\rm HK}$$>$$-$5.0 and increases with the activity level of the star (Fig.~\ref{EW_sensitivity}-bottom). Using the log R$^\prime_{\rm HK}$-age relation calibrated by \citet{Lorenzo-Oliveira18} on solar twin stars, we conclude that spectra of Sun-like stars younger than 4-5 Gyr are likely affected by this phenomenon.

\item The observed dependence of line EWs from chromospheric activity can be ascribed to the Zeeman broadening of absorption lines that form near the top of the stellar photosphere where magnetic fields are stronger or to the presence of cold stellar spots that can increase the EWs of lines at low energy potentials.

\item Stellar parameters inferred from the simultaneous search for three spectroscopic equilibria of iron lines (i.e., excitation equilibrium, ionisation balance, and the relation between log N$_{FeI}$ and the reduced EWs) are also influenced by the intensification of absorption lines due to stellar activity (Fig. \ref{sensitivity_parameters_MCMC} and \ref{MCMC_prediction}). Namely, the EW increase of strong Fe I lines leads to higher $\xi$ values, which, as a consequence, lowers [Fe/H]. This effect is visible for stars with log R$^\prime_{\rm HK}$$>$$-$5.0 (i.e., stars younger than 4-5 Gyr) and increases for more active stars. We also observe a decrease of T$_{\rm eff}$ with chromospheric activity, which is evident for stars with log R$^\prime_{\rm HK}$$>$$-$4.5 (i.e., younger than 1 Gyr). These effects are not negligible at the typical precision of large spectroscopic surveys. From our analysis, we have not observed any variation in log~g.

\item The intensification of absorption line due to chromospheric activity and the consequent rise of $\xi$ has the effect of changing the abundances of specific elements. We have identified three classes of elements.  The first class includes the species such as Si, Sc, Ti, Mn, Fe, Ni, and Cu, that are detected mostly through medium-strong lines (EW$\sim$40-70~m$\rm \AA$) and that formed deep in the stellar photosphere ($\tau_\lambda>$1). These lines are not significantly intensified by magnetic fields or stellar spots, but they are very sensitive to $\xi$. Therefore, the growth of their EWs is not enough to compensate for the rise of $\xi$. For this reason, the abundance of these elements decreases as a function of the stellar activity index. The second class of elements includes C, Na, Al, S, V, Co, Y, and Zr. The lines of these elements are very weak and formed deep in the atmosphere. They are nor affected by stellar activity, nor sensitive to $\xi$. Therefore, the change of their abundance is nearly zero. Finally, the elements Mg, Ca, Cr, and Ba are detected through strong lines that formed in the upper atmosphere. These lines are both intensified by stellar activity and sensitive to $\xi$, in a way that one effect counterbalance the other. Therefore the change in their abundances is also nearly zero. We stress again that these conclusions depend on the line list that is employed in the spectroscopic analysis.

\item The finding that stellar parameters and abundances can vary as a function of the stellar activity level has several fundamental implications on different topics in astrophysics. For example, studies of Galactic Chemical Evolution will have to consider that the effect described in this paper can artificially affect the chemical abundances obtained through spectroscopic analysis of quantities that depend on the stellar activity, which also scales with the stellar age. On the other hand, the modulation that stellar activity induces on the EWs of the strongest lines (such as the Ba line at 5853 $\rm \AA$, see Fig.~\ref{EW_sensitivity}) can be used to trace the phase of the activity cycle in addition to the Mt. Wilson S-index. This can result particularly useful for planet detection techniques. 
\end{itemize}
 

With this work we aim at improving the current techniques of spectroscopic analysis by highlighting the limitations and inconsistencies caused by the simplistic assumption that stellar spectra are not affected by magnetic fields and stellar spots. It is central to our progress in different areas of astrophysics that we overcome this difficulty. Therefore, the next step of this research will necessarily be the identification of the main cause(s) of the phenomenon described above (e.g., magnetic fields and/or stellar spots) which will then allow us to find a definitive solution to these limitations (e.g., a new method of spectroscopic analysis and stellar models that incorporate the effects of magnetic fields and stellar spots). 

However, the undesired effects of magnetic activity on the spectroscopic analysis can be already hindered by a strategic choice of absorption features in the master list (e.g., see \citealt{Baratella20}). From our study it is clear that the sensitivity of the stellar parameters to the activity index is mainly due to the use of many medium and strong Fe lines formed in the upper layers of the stellar atmosphere. Namely, out of the 95 Fe~I lines in our master list (Table~\ref{linelist}), 11 have EWs measured in the Solar spectrum that are within 70-80 m$\AA$, nine fall in the range of 80-90 m$\AA$, and six have EWs $\geq$100 m$\AA$. On the other hand, \citet{Galarza19} showed that by choosing a list of weaker Fe lines non sensitive to variations to chromospheric activity they have been able to obtain smaller microturbulences and statistical errors in the spectroscopic analysis of the young solar twin HD~59976. Therefore, a partial solution to the activity problem would be to employ a list of weak Fe lines to determine the stellar parameters, such as the line list used by \citet{Nissen15} which includes only Fe lines weaker than 70 m$\AA$. However, an indiscriminate choice of only the weakest Fe lines could also significantly reduce the number of lines at low excitation potential and high reduced EW, necessary to reach the excitation/ionisation equilibria. Alternatively, \citet{Baratella20} proposed a new method based on titanium lines to derive the spectroscopic surface gravity, and most importantly, the microturbulence parameter, while a combination of Ti and Fe lines is  used to obtain effective temperatures.

The use of weak lines can certainly improve the abundance determination of individual elements. However, the problem remains for those element, such as Ba, that are observed only through strong lines formed in the upper layers of the stellar atmosphere. In these cases, the only viable solution could come from magneto-hydrodynamical modelling of stellar atmospheres.

\acknowledgments
 LS and AIK acknowledge financial support from the Australian Research Council (Discovery Project 170100521). ARC thanks the support from the Australian Research Council (DECRA 190100656). JM thanks support by FAPESP (2018/04055-8) and CNPq (Bolsa de Produtividade). JYG acknowledges the support from CNPq. Finally, the authors gratefully acknowledge the constructive comments offered by the anonymous referee.

%

\vspace{5mm}
\facilities{ESO 3.6m - La Silla Observatory, HARPS \citep{mayor03}.}


\software{ \texttt{qoyllur-quipu} \citep{Ramirez14}, \texttt{MOOG} \citep{Sneden73}, \texttt{{\tt pymc3}} \citep{Salvatier16}, \texttt{Stellar diff} (\url{https://github.com/andycasey/stellardiff}).}


\bibliography{Bibliography} 



\end{document}